# Outline of a unified Darwinian evolutionary theory for physical and biological systems

Carlos Baladrón

Departamento de Física Teórica, Atómica y Óptica, Universidad de Valladolid, 47071 Valladolid, Spain. Email: baladron@cpd.uva.es

and

Andrei Khrennikov

International Center for Mathematical Modeling in Physics and Cognitive Science, Linnaeus University, Växjö, Sweden. Email: andrei.khrennikov@lnu.se

Abstract: The scheme of a unified Darwinian evolutionary theory for physical and biological systems is described. Every physical system is methodologically endowed with a classical information processor what turns every system into an agent being also susceptible to evolution. Biological systems retain this structure as natural extensions of physical systems from which they are built up. Optimization of information flows turns out to be the key element to study the possible emergence of quantum behavior and the unified Darwinian description of physical and biological systems. The Darwinian natural selection scheme is completed by the Lamarckian component in the form of the anticipation of states of surrounding bio-physical systems.

# 1.Introduction.

There are certain frameworks, like universal Darwinism (Dawkins,1983) or generalized Darwinism (Aldrich, 2008), that apply the essence of Darwinism (evolution under natural selection of systems possessing the properties of variation, selection and retention) to different domains of knowledge. Darwinism has been also applied in physics as a possible mechanism explaining different fundamental problems (Smolin, 2006, Zurek, 2009). In this article a theory in progress (Baladrón, 2010, 2014; Baladrón and Khrennikov, 2016) is reviewed whose aim is to explore the possibility of unifying Darwinism for the physical and the biological realms. In a certain sense, it can be considered an extension of Darwinism to the physical domain, but at the same time, in a deeper level, a unification of the physical and biological descriptions under a generalized Darwinian perspective in which information would play a central role applied to the specialized physical and biological domains. In the end, biological systems are built with matter. Therefore, biological systems must also comply with physical laws, and a connection between matter and life is already established in the physical realm. However, might there be a deeper connection? The function of life is to continue to exist, in a constantly changing environment (Eigen, 2013; Hill and Nowak, 2014). Could this be also asserted for the function of matter from the perspective of quantum mechanics? Thus, the question arises as to whether Darwinism could be a unifying approach for the description of matter and life under the overarching framework of information.

This theory studies the possibility of generating implementing quantum-like behavior in classical dynamical physical systems that are methodologically endowed with a probabilistic classical Turing machine (see Encyclopedia of Mathematics, 2013. Turing machine), i.e. basically an information processor plus a random number generator –for biological systems that will be explicitly considered in Section 3 the information processor would be the network of the processors associated to their physical constituents.[1] Every physical system would then be governed by a program that would have been developed under the action of Darwinian evolution starting at the Big Bang ($t = 0$) in the physical space and a corresponding initial information state of minimal content coding the initial conditions. These initial conditions in the information space would be the following. Initial state and algorithm defining the random number generator; the basic set of abstract elements and operations that defines the Turing machine (see Barker-Plummer, 2016; and Section 2.1); the connecting rules between the information space and the physical space of the bare system; and finally at $t = 0$ the blank state for the information about the physical space (surrounding systems) from an initial information blank state. In this way every fundamental system would become an agent in an evolutionary scenario in which there would be no universal laws, but systems driven by programs that would evolve submitted to natural selection pressure (see Fig. 1 for a representation of the self-interaction process in a system resulting from the interplay between the bare material system and the probabilistic classical Turing machine). This information-theoretic Darwinian scenario would provide meaning to the information conveyed by every agent. The meaning of information would be its utility for the stability or survival of the agent. Quantum-like behavior would then result from the optimization of past, present and anticipated classical information flows for the stability of the system. Therefore, this theory might shed new light (Baladrón and Khrennikov, 2016) on the concept of quantum information (Nielsen and Chuang, 2000; Jaeger, 2007) and its relation with classical information.

---

[1] In animas this probabilistic Turing machine is physically based on the nervous system; for plants this is the system of signaling between cells representing a kind of distributed cognition; for individual cells, the Turing machine like computational device is composed of epigenome and the network of molecular signaling. The quantum-like

In our model, bio-physical systems demonstrate the ability for anticipation of states of surrounding systems. This ability is the output of functioning of Turing machines associated with systems.[2]

---

signatures of network functioning were discussed in De Barros and Suppes (2009), and De Barros (2012).

[2] The anticipation component of the evolutionary dynamics can be compared with the active information field interpretation of the wave function, see Bohm and Hiley (1993). Such a component also plays the fundamental role in theory of partially directed evolution which was developed in the works Melkikh (2014), and Melkikh and Khrennikov (2015, 2016) and it was based on exploring the mathematical formalism of quantum information theory. This anticipation dimension can be treated as the Lamarckian element of the model, see Asano et al. (2014, 2015) for extended discussion.

This information-theoretic physical Darwinism, which would explain the emergence of quantum behavior and would underlie the unification of the evolutionary description of physical and biological systems, would represent a solution to the conundrum of the meta-laws (Unger and Smolin, 2015), which is intrinsically associated to the analysis of the possibility of changing laws in cosmology (Unger and Smolin, 2015), since once the possible variability of the laws is admitted, then this mutability could also affect the manner in which the natural laws change –i.e., the meta-law--, and the problem of explaining the historical change of laws would have just been transferred to a higher level (the conundrum of the meta-laws). The proposed information-theoretic physical Darwinism would solve this problem acting as a natural self-generated meta-law, therefore eliminating the necessity of a recurrent explanation to higher levels.

The interaction between the world of ideas and the material world –or in a more physical terminology between information and matter—has been present in the history of physics from the Ancient Greece (Pombo, 2010, 2015). Since then the connections and interplay between these two worlds have been analyzed by certain schools of thought as a possible alternative key element in the understanding of nature.

The article is organized as follows. The theory and model for a fundamental physical system is analyzed in Section 2. The model for a biological system is described and the results discussed in Section 3. Finally, the conclusions are drawn in Section 4. The paper also contains the extended appendix devoted to interconnection of our evolutionary model with Whiteheadian metaphysics.

**2. Information-theoretic Darwinian model for a fundamental physical system.**

An information-theoretic Darwinian approach (Baladrón, 2010, 2014; Baladrón and Khrennikov, 2016) applied to classical physical systems is presented in this Section. It is expected to explain the emergence of quantum mechanical behavior. Darwinism is going to play the role of the self-generated meta-law that determines the evolution of the programs – which in turn govern the behavior of the fundamental physical systems-- stored on the probabilistic classical Turing machines from the initial informational blank state –corresponding to the Big Bang in the physical space— in which the physical systems are under the control of their respective randomizers.

The two main characteristics of this approach are influenced by two famous sayings of Wheeler. First, the weight of information as a fundamental element shaping matter behavior is deeply swayed by Wheeler's dictum "*it from bit*" (Wheeler,1990). Second, an appealing answer to the challenging Wheeler's question "*why the quantum?*" is tentatively supplied by the usage of Darwinism in order to explore the possibility that quantumness might emerge from classicality as an efficient solution leading to steady physical systems.  In other words, if the world is going to present regularities, then it must be quantum.

## 2.1 Composition of a physical system

A fundamental physical system is defined in this theory by its continuous trajectory in physical space $X(t)$ and its average mass $m$. Additionally, every system is complemented with both a methodological classical Turing machine[3] (see Barker-Plummer, 2016 Encyclopedia of Mathematics, 2013. Turing machine) and a random number generator defined on an information space. The randomizers completely control the self-interaction process of the physical systems at the first stage of the dynamics. As the information content increases with time, the programs progressively develop and take control of the system, although the randomizers will continue playing a role so as to optimize the strategies of the systems (Ross, 2006). Thus, every physical system can be considered an agent that --as an object-- obeys the classical physical constraints of conservation of momentum and energy when receiving the carriers of momentum, energy and information emitted by the surrounding systems, and --as a subject-- emits a self-interacting self-interaction carrier[4] --of a certain momentum and energy-- being the output of the program that rules the behavior of the system.

---

[3] In a simplified manner, a Turing machine can be defined as an abstract machine that is at any time in a state out of a list of specified states and is constituted by an one-dimensional infinite tape of adjacent cells in which information is or can be stored (in its simplest way as a string of binary code: zeros and ones, a symbol at every cell in the tape); a read/write head --located on a single cell-- that can either write a symbol ("0" or "1") on the cell or move left or right along the tape, one cell at most at every step of the computation; and finally the program, i.e. a table of transition rules, every rule containing four elements: the machine's current state, a symbol ("0" or "1"), the machine's next state, and the action to be performed by the head. Every rule establishes that if the current state of the machine and the written value of the cell on which the head is located are the two first elements in the transition rule, then the next state of the machine and the action to be performed by the head are the third and fourth elements in the transition rule. An interpretation of the string of zeros and ones written on the tape when the machine halts must be added in order to give a meaning to the computation. In our case, the result of the computation would contain the parameters of the carrier to be emitted by the physical system.

The state of a system is thus characterized by its position $X$ in the physical space – as in Bohmian mechanics (Goldstein, 2016)—, its random number generator or randomizer $R$, and its program $P$, the latter two elements defined on the information space. Hence, the theory possesses a minimalist realistic ontology.

A probabilistic classical Turing machine is able to do the very same tasks as a quantum Turing machine (Timpson, 2007), in particular, the simulation of quantum behavior for a physical system. The randomizer is necessary to calculate the probabilities for the final states in the probabilistic classical Turing machine. The difference between both types of Turing machines resides in the efficiency. There are certain problems that a quantum Turing machine would solve much more efficiently. An example is the Shor algorithm (Shor, 1997) for factoring numbers on a quantum computer that has not known parallel counterpart on a classical computer. Thus, if it is expected to generate quantum behavior in real time by means of a probabilistic classical Turing machine, then it is necessary to implement a procedure developing efficient enough algorithms. This theory contends that Darwinian natural selection might be able to accomplish such a task.

Therefore, the backdrop of the theory is depicted by a physical space populated by systems with information processing capabilities that are governed by software programs. The degree of adaptation of every possible program to a certain environment can be measured by a merit function that defines a landscape of possible algorithms (Wolfram, 2002) –in evolutionary biology, the concept of fitness landscape was introduced by Wright (1932)--. So, assuming variability on the information space, e.g. through the process of read/write operation in the Turing machine, the program could move around the abstract landscape of possible algorithms and the natural selection pressure would shape the evolution of the population of systems in physical space towards the conditional peaks in the landscape on information space.

---

[4] The interaction carriers are supposed to represent the basic mode of interaction between fundamental physical systems that should give account, as a result of Darwinian evolution, of the present-known interactions (through gauge bosons) between matter particles when quantum equilibrium had been reached.

The physical system so defined is then characterized as a generalized Darwinian system (Aldrich et al., 2008) with the properties of variation, selection, and retention –i.e. the fit variants are kept in the new version of the program written on the tape of the Turing machine--. The population of the best adapted systems upon random variations on the information space would increase –i.e. selection—and the new features of these fittest systems would be retained in the new versions of the program written on the tape of every corresponding Turing machine –i.e. retention--.

**2.2 Dynamics of a physical system: Bit from it and it from bit**

The probabilistic classical Turing machine at $t = 0$ only contains the randomizer, the basic set of elements and operations that defines the machine (see Encyclopedia of Mathematics, 2013. Turing machine), and the rules of connection between the information space on which the Turing machine operates and the physical space of the bare system.

These rules state the following. First, the information about the physical space conveyed by the carriers that hit the system is stored on the tape of the Turing machine. And second, this information in the short term serves as input for the controlling program (transition rules) whose output determines the self-interaction of the system.

At $t = 0$ the emission of momentum-energy carriers is entirely at random, since there is no program commanding the system, only the randomizer. But as time increases the information accumulates and a program that takes control of the system evolves under natural selection.

In this theory every system is both an emitter and a receiver of information. This information is about the position that, at any time, every system occupies in the physical space. The carriers emitted by a system convey information about the position of the system (see Fig.1 for a graphical representation of a self-interaction cycle of a system). Then a receiver can construct –as in a radar problem-- a probability distribution function $\rho_i(t)$ for the position of every surrounding system, $i$, from the information transported by the impinging carriers.

Symbolically, it might be said that every system as an emitter generates information about the real positions (paraphrasing Wheeler: *it*, i.e. the property that defines matter in the physical space in the theory) occupied by the system. This information (paraphrasing Wheeler again: *bit*) transported by the emitted carriers enables the receiver –by means of the information processor-- to estimate a probability distribution function (*bit*) for the position of the emitters $\rho_i(t)$, where $i$ denotes every emitter. A convenient way of measuring the amount of information contained in that distribution function is the Fisher information measure $I$ (Frieden, 1989; Honig, 2009) that is related to Shannon information, but better adapted to mathematically deduce a differential equation describing a dynamical process.

If the algorithm stored on the information processor were was complex enough, then it might could further elaborate the information contained in the probability distribution function about the current configuration of the surrounding systems, and estimate by means of an anticipation software module $A$ the possible future position configuration for the outside systems at time $t + \Delta t$. But this anticipations or projections of the future positions of the surrounding systems are nothing but the *subjective beliefs* $\psi(t + \Delta t)$--in the terminology of QBism, see Healey (2016)-- of the system about the expected behavior of the outside observed systems at time $t + \Delta t$. As in Bohmian mechanics (Goldstein, 2016), in this theory all the magnitudes must be characterized in terms of systems configurations (positions), since position is the defining property of a system.

At this point, the meaning of information appears connected with the relevance or utility of the information for the stability or, in biological terms, survival of the system. The key point is the capability of anticipation through the processing of information that allows the system to assign meaning to the different consequences of an action in relation with the stability of the system and considering the estimated future configurations of the environment.

A final step has to be performed; it is to determine the output of the program, i.e. the self-interaction –the parameters of the carrier to be emitted for optimizing the stability of the system--. The result of the execution of the program after the elaboration of information (*bit*) is the new position (*it*) of the system.

A coherent series of interconnected definitions of information properties --resembling the *pragmatic information* project (Gernert, 2006; Roederer, 2016) and the concept of *active information* in Bohmian mechanics (Bohm and Hiley, 1993; Gernert, 2006; Hiley, 2002)— for a microscopic physical agent has been established. In the next paragraphs, it is going to be discussed how this theory could drive these generalized Darwinian physical systems to behave quantum-like.

## 2.3 Proposal for an evolutionarily stable strategy (ESS)

In classical evolutionary biology the fitness landscape for possible strategies is considered static. Therefore optimization theory is the usual tool in order to analyze the evolution of strategies (Nowak and Sigmund, 2004) that consequently tend to climb the peaks of the static landscape. However in more realistic scenarios the evolution of populations modifies the environment so that the fitness landscape becomes dynamic. In other words, the maxima of the fitness landscape depend on the number of specimens that adopt every strategy (frequency-dependent landscape). In this case, when the evolution depends on agents' actions, game theory is the adequate mathematical tool to describe the process (Nowak and Sigmund, 2004). But this is precisely the scheme in the present study, namely, the evolving physical laws (i.e. algorithms or strategies) are generated from the agent-agent interactions (bottom-up process) submitted to natural selection.

The concept of evolutionarily stable strategy (ESS) (Maynard Smith, 1974) is central to evolutionary game theory. An ESS is defined as that strategy that cannot be displaced by any alternative strategy when being followed by the great majority –almost all—of systems in a population. In general, an ESS is not necessarily optimal; however it might be assumed that in the last stages of evolution --before achieving the quantum equilibrium-- the fitness landscape of possible strategies could be considered static or at least slow varying. In this simplified case an ESS would be one with the highest payoff therefore satisfying an optimizing criterion. Different ESSs could exist in other regions of the fitness landscape.

An evolutionarily stable strategy (ESS) (Maynard Smith, 1974) is defined as that strategy that cannot be displaced by any alternative strategy when being followed by the great majority –almost all—of systems in a population. Regarding the simplified case of a static landscape of strategies (Nowak and Sigmund, 2004) an ESS would be one with the highest payoff therefore satisfying an optimizing criterion. In the information-theoretic Darwinian approach it seems plausible to assume as optimization criterion the optimization of information flows for the system. Following this criterion a set of three regulating principles (Baladrón, 2014; Baladrón and Khrennikov, 2016) is proposed:

*Principle 1 (Structure): The complexity of the system is optimized (maximized).*

The definition that is adopted for complexity in this theory is Bennett's logical depth (Bennett, 1988) that for a binary string is the time needed to execute the minimal program that generates such string[5]. Then the complexity of a system at time $t$ in this theory would be the Bennett's logical depth of the program stored at time $t$ in its Turing machine.

The increase of complexity is a characteristic of Lamarckian evolution, and it is also admitted (Adami, 2003; Miconi, 2008) that the trend of evolution in the Darwinian theory is in the direction in which complexity grows, although whether this tendency depends on the timescale --or some other factors-- is still under discussion.

---

[5] There is no a general acceptance of the definition of complexity, neither is there a consensus on the relation between the increase of complexity –for a certain definition— and Darwinian evolution. However, it seems that there is some agreement on the fact that, in the long term, Darwinian evolution should drive to an increase in complexity in the biological realm for an adequate natural definition of this concept --see Baladrón and Khrennikov (2016) for a discussion on the connections among some definitions of complexity--. Bennett's logical depth, in principle, seems to be a satisfactory characterization of complexity that catches the essentials of biological evolution (Deutsch, 1985)

*Principle 2 (Dynamics): The information outflow of the system is optimized (minimized).*

The information is the Fisher information measure *I* (Frieden, 1989; Honig, 2009) for the probability density function of the position of the system.

According to Frank (2009), natural selection acts maximizing the Fisher information within a Darwinian system. As a consequence, assuming that the flow of information between a system and its surroundings can be modeled as a zero-sum game (Frieden and Soffer, 1995), Darwinian systems would follow the Principle 2.

*Principle 3 (Interaction): The interaction between two subsystems optimizes (maximizes) the complexity of the total system.*

The complexity is again equated to the Bennett's logical depth.

The role of *Principle 3* is central in the generation of composite systems, therefore in the structure for the information processor of composite systems resulting from the logical interconnections among the processors of the constituents. In Section 3, this question will be further discussed.

There is an enticing option of defining the complexity of a system in contextual terms as the capacity of a system for anticipating the behavior at $t + \Delta t$ of the surrounding systems included in the sphere of radius $r$ centered in the position $X(t)$ occupied by the system (Baladrón and Khrennikov, 2016). This definition would directly drive to the maximization of the predictive power for the systems that maximized their complexity. However this magnitude would definitely be very difficult to even estimate, in principle much more than the usual definitions for complexity --see Baladrón and Khrennikov (2016) for a commentary on other definitions of complexity in the perspective of this theory--.

## 2.4 Quantum behavior

Quantum behavior of microscopic systems should now emerge from the proposed ESS. In other terms, the postulates of quantum mechanics should be deduced from the application of the three regulating principles on our physical systems endowed with an information processor.

Let us apply the *Principle 1*. It is reasonable to consider that the maximization of the complexity of a system would in turn maximize the predictive power of such system. And this optimal statistical inference capacity would plausibly induce the complex Hilbert space structure for the system's space of states. This result has been analyzed in several studies (Aerts , 2008; De Raedt et al., 2013, 2015; Summhammer, 1994, 2007). See also Baladrón and Khrennikov (2016) and references therein.

Let us consider the *Principle 2*. This is basically the application of the principle of minimum Fisher information or maximum Cramer-Rao bound (Frieden, 1989) on the probability distribution function for the position of the system. Frieden (1989) derives the Schrödinger equation from this hypothesis.

The concept of entanglement (Bub, 2016) seems to be determinant to study the generation of composite systems, in particular in this theory through applying the *Principle 3*. The theory admits a simple model that characterizes the entanglement between two subsystems as the mutual exchange of randomizers $(R_1, R_2)$, programs $(P_1, P_2)$ –with their respective anticipation modules $(A_1, A_2)$-- and wave functions $(\psi_1, \psi_2)$. In this way, both subsystems can anticipate not only the behavior of their corresponding surrounding systems, but also that of the environment of its partner entangled subsystem (Baladrón, 2016) –see the Appendix for a deepest discussion on anticipation in the present theory--. In addition, entanglement can be considered a natural phenomenon in this theory, a consequence of the tendency to increase the complexity, and therefore, in a certain sense, an experimental support to the theory.

There are other two three studies that might sustain support the information-theoretic Darwinian approach to quantum mechanics. First, certain quantum-like properties obtained in the laboratory for a macroscopic liquid drop that bouncing on a vibrating bath forms a dynamical system with the surface waves that it produces (Perrard et al., 2014, 2016). They These authors show that the liquid drop, which is piloted by the surface waves generated by the drop itself, can be formally described by a Turing machine (Perrard et al., 2016) being the surface waves the locus of the information. Second, the study of Chatterjee et al. (2013) that shows in computer simulations that evolution through certain mechanisms can perform in polynomial time certain tasks that most evolutionary processes take exponential time to perform. Third, the astronomical and cosmological observations should constitute in a near future a fundamental arena in which this theory could be confronted (Baladrón, 2014). The proposal of Valentini (2007) to test Bohm-like theories (deterministic hidden variables theories) by searching anomalies in astronomical observations which could correspond to the presence of remnants of quantum non-equilibrium in certain regions of the universe might be adapted to the present information-theoretic Darwinian approach.

In addition, the information-theoretic Darwinian approach is a minimalist realist theory –every system follows a continuous trajectory in time, as in Bohmian mechanics--, a local theory in physical space –in this theory apparent nonlocality, as in Bell's inequality violations (Khrennikov, 2016), would be an artifact of the anticipation module in the information space--, although randomness would necessarily be intrinsic to nature through the random number generator methodologically associated with every fundamental system at $t = o$, and as essential ingredient to start and fuel – through variation— Darwinian evolution. As time increases, random events determined by the random number generators would progressively be replaced by causal events determined by the evolving programs that gradually take control of the elementary systems. Randomness would be displaced by causality as physical Darwinian evolution gave rise to the quantum equilibrium regime, but not completely, since randomness would play a crucial role in the optimization of strategies –thus, of information flows—as game theory states (Ross, 2006).

In summary, at the price of assuming fundamental systems endowed with a random number generator and a classical information processor, quantum mechanics might be explained through the action of Darwinian evolution in terms of a minimalist realist, local and quasi-causal theory –quasi-causal meaning that although the theory is intrinsically and fundamentally random, however it is random with a cause, that of developing an optimal strategy for the stability of the system.

### 3. Information-theoretic Darwinian model for a biological system

A natural extension of the information-theoretic Darwinian approach for biological systems is obtained taking into account that biological systems are constituted in their fundamental level by physical systems. Therefore it is through the interaction among physical elementary systems that the biological level is reached after increasing several orders of magnitude the size of the system and only for certain associations of molecules – biochemistry.

In particular, this viewpoint lies in the foundation of the *"quantum brain"* project established by Hameroff (1994) and Penrose (1989, 1994). They tried to lift quantum physical processes associated with microsystems composing the brain to the level of consciousness. Microtubulas were considered as the basic quantum information processors. This project as well the general project of reduction of biology to quantum physics has its strong and weak sides, see, for example, (Tegmark, 2000) for discussion. One of the main problems is that *decoherence* should quickly wash out the quantum features such as superposition and entanglement. (Hameroff and Penrose would disagree with this statement. They try to develop models of hot and macroscopic brain preserving quantum features of its elementary micro-components.)

However, even if we assume that microscopic quantum physical behavior disappears with increasing size and number of atoms due to decoherence, it seems that the basic quantum features of information processing can survive in macroscopic biological systems (operating on temporal and spatial scales which are essentially different from the scales of the quantum micro-world). The associated information processor for the mesoscopic or macroscopic biological system would be a network of increasing complexity formed by the elementary probabilistic classical Turing machines of the constituents. Such composed network of processors can exhibit special behavioral signatures which are similar to quantum ones, see (De Barros and Suppes, 2009; De Barros, 2012). We call such biological systems *quantum-like.* In the series of works Asano et al. (2014, 2015), there was developed an advanced formalism for modeling of behavior of quantum-like systems based on theory of open quantum systems and more general theory of adaptive quantum systems. This formalism is known as *quantum bioinformatics.*

The present quantum-like model of biological behavior is of *the operational type* (as well as the standard quantum mechanical model endowed with the Copenhagen interpretation). It cannot explain physical and biological processes behind the quantum-like information processing. Clarification of the origin of quantum-like biological behavior is related, in particular, to understanding of the nature of entanglement and its role in the process of interaction and cooperation in physical and biological systems.

We remark that qualitatively the information-theoretic Darwinian approach supplies an interesting possibility of explaining the generation of quantum-like information processors in biological systems. Hence, it can serve as the bio-physical background for quantum bioinformatics. There is an intriguing point in the fact that if the information-theoretic Darwinian approach is right, then it would be possible to produce quantum information from *optimal flows of past, present and anticipated classical information* in any classical information processor endowed with a complex enough program. Thus the unified evolutionary theory would supply a physical basis to QIB.

## 4. Conclusions.

A unified coherent description for physical and biological systems in terms of Darwinian evolution (completed by the anticipating component of the Lamarckian type) might be possible considering the results discussed in this article. Endowing classical elementary physical systems with an information processor and a randomizer transforms these systems in microscopic agents what allows exploring the possibility of generating quantum behavior from the optimization of information flows through Darwinian evolution. The possible emergence of quantum mechanics as an ESS has been discussed. Thus, nature --that is considered intrinsically random-- is described in terms of minimalist realist systems that interact locally in physical space; apparent nonlocality being the result of anticipated possible information elaborated in the information space of every system from past and current information conveyed by interacting carriers among systems. This theory also enables a natural, continuous transition from physical to biological systems. In addition, certain difficulties of biology admit a natural explanation in this unified scheme in which randomness, information and anticipation are crucial concepts. A complete mathematical theory of evolution is a difficult task for the future, but, as this study suggests, it might be central for the advance in the knowledge of nature.

# Appendix: Whiteheadian perspective on bio-physical evolutionary model

Whitehead (1929, 1933) was one of the first philosophers who created a new philosophic system under the influence of new born quantum theory. His philosophic system is known as *Philosophy of Organism.* It played the important role in philosophic rethinking of foundations of physics, see, e.g., Shimony (1965). The most striking feature of Philosophy of Organism is that it was the first attempt of unification of the physical and mental worlds on the basis of *protomental elements.* They can be considered as quanta of mentality and at the same time as basic elements of the physical world (treated as the world which is composed of quanta).

According to Whitehead(1929, 1933), the world consists of events. The basic events are so-called *actual occasions*. Objects arise as stable patterns in chains of actual occasions. An actual occasion is meaningful only with respect to a chain of the antecedent occasions. The basic features of a new-coming actual occasion are anticipated in antecedent actual occasions. However, the process of creation of new actual occasions is not deterministic. There is always present an element of random creation, creation of features which were not present in antecedent actual occasions. Such "random generator" associated with the process of creation is the main source of novelty in the world.

Let us consider the process of creation of an actual occasion $Q$ in more detail. Denote this process by the symbol $\pi$. In the simplest case, we have a chain of actual occasions:

$$\ldots \varPhi,\ldots,\varSigma,\ldots,\varLambda,Q,\ldots$$

For example, such a chain can represent some enduring object – a persistent feature of the occasions in the chain. The process of creation of *Q* is based on memory about all antecedent occasions,…*Φ*,...,*Σ* ,..., *Λ*. However, the use of the memory resource is just one side of the process of design of *Q*. In some sense, this is still pure physical counter part of π. Now we consider a more striking side of π. At the step of creation of *Q,* the process π interacts with other actual occasions "nearby" *Q* . In general these occasions need not belong to the chain …*Φ*,...,*Σ* ,..., *Λ*. Moreover, π interacts with other processes of creation of new occasions. By collecting information about such neighboring occasions and processes, π can anticipate the future outputs of the latter (of course, such anticipation is probabilistic). This information plays the important role in creation of *Q*. [6] The anticipatory feature of the process of creation of new actual occasions can be treated as a mental counterpart of this process.

Thus, for Whitehead (1929, 1933), even entities which are typically treated as of the purely physical nature, in fact, have some mental features. This is the natural basis for the unified description of physical and biological processes. This dimension of Philosophy of Organism is very supporting for our grand unification project for bio-physical evolution.

Finally, by overviewing the basics of Philosophy of Organism we present the Whiteheadian viewpoint on the appearance and evolution (!) of physical laws. Consider a society (ensemble)[7] of actual occasions. There may be characteristics common to all or practically all members of society. Hence, the prehensions of new occasions will be uniform in some respects and such common characteristics would have the *tendency* to persist. We emphasize that this is not just the memory based persisting. This is also the common anticipation based persisting. Physical laws are considered as special class of such commonality in prehensions.

---

[6] We remark that in his Philosophy of Organism Whitehead did not use the informational paradigm. Here we started to use it. On one hand, this simplifies the presentation of Whitehead's theory. On the other hand, it is useful as preparation to mapping of the basic entities of Philosophy of Organism to our quantum-like model of bio-physical evolution. We also remark that the closeness in the spaces of occasions and processes can also be formalized by using the informational paradigm, as a topological structure on the *information space*, see, e.g., (Khrennikov, 1999).

Thus according to Philosophy of Organism physical laws are outputs of evolution of societies of actual occasions. In particular, they are not "given and firmly established" by nature. They are created in the processes of evolution based on the process of creation of actual occasions from antecedent actual occasions. Physical laws are not forever. They can be modified in accordance with modification of common characteristics in societies of actual occasions. Another important characteristic of physical laws (in the Whiteheadian framework) is that they are not deterministic. Since processes of creation of new actual occasions contain "elements of freedom", even in homogeneous societies of occasions the processes of creation are random.

We also stress that physical laws have no absolute validity. They hold for special societies of occasions. Moreover, the degree of their probabilistic validity can also depend on society. We also point to a very interesting Whitehead's idea about possible "degradation" of laws of physics. The degree (probability) of validity of some law can decrease in the process of evolution of a society of actual occasions. Of course, the same picture is valid for biological and social laws. In short, *for Whitehead physical, biological, and social laws are products of evolution*.

---

[7] Whitehead operated with the notion "society" and we follow him, although it does not match physical terminology. It may be more natural to speak about ensembles.

We now try to match the basic elements of Philosophy of Organism with our quantum-like model of bio-physical Darwinian evolution. The latter is based on the quantum information perspective. Therefore it is fruitful to consider Whitehead's philosophy of bio-physical reality from the same (quantum informational) perspective. Thus we can speak not about the mental (anticipating) component of each actual occasion, but about its information component.

The main difference between Whitehedian and our approaches is that the basic notion of our grand unification model is the notion of a system, say *S*. A process appears as a chain of state transformations for *S*,

$$...\rightarrow...\rightarrow \sigma \rightarrow \sigma' \rightarrow \sigma'' \rightarrow... \rightarrow... \qquad (W1)$$

However, in the information approach the system is merely an information transformer. We are interested in *S not* as an object, but just as the symbolic representation of the process (W1). For Whitehead, elementary quantum particles are just special chains of actual occasions having some enduring features which characterize elementary particles.

By Philosophy of Organism each actual occasion has the finite duration in time and it covers an extended domain in space, i.e., it cannot be modelled as a point-wise structure. This is one of the most cardinal differences from classical physics.

We first analyze the finite time duration of an actual occasion. In ESS and more generally in our bio-physical Darwinian evolution theory, the Whiteheadian creation of an actual occasion during some finite time interval can be associated with a step of processing done by the Turing machine which endows a quantum physical system or a biological system exhibiting quantum-like behavior.

Both Whitehead (1929, 1933) and the authors (Baladrón, 2010, 2016; Baladrón and Khrennikov, 2016) assumed the presence of a random element in the evolutionary dynamics. In our model, each information processor is endowed with a random generator, as a part of its probabilistic Turing machine. In Whitehead's Philosophy of organism the presence of a random element in creation of new actual occasions is the intrinsic feature of nature. It creates permanently novel characteristics of actual occasions.

The memory about preceding actual occasions is represented in the internal state of the Turing machine. Each Turing machine also receives information about previous and present states of "neighboring" Turing machines. This information is used to anticipate the coming outputs of these machines and produce the own output adapted not only to the present states of surrounding information processors, but even to their future outputs. Of course, this adaptation-anticipation is of the probabilistic nature.

Spatial extension of an actual occasion $Q$ can be coupled with two fundamental quantum structures, *superposition and entanglement*.

In the Whiteheadian metaphysical model, each individual actual occasion is spatially extended, since its creation is based on a chain of antecedent occasions and anticipation of appearance of surrounding occasions.

First we consider the process of memory induced spatial extension of actual occasions. Even if the "initial actual occasion" $Q(0)$ was pointwise (represented by a point in physical space $\mathbf{R}^3$), it will diffuse and cover a domain in $\mathbf{R}^3$. A few generations of pointwise actual occasions, $Q(0)$, $Q(t_1)$,..., $Q(t_n)$, generate a spatially extended actual occasion. It will be concentrated in the spatial domain with the skeleton $X(t_1)$, ..., $X(t_n)$, where $X(t_j)$ (belonging $\mathbf{R}^3$) is the vector of coordinates of the (pointwise) actual occasion $Q(t_j)$. Already the memory about say two preceding pointwise actual occasions $Q(t_1)$ and $Q(t_2)$ with the coordinates $X(t_1)$, $X(t_2)$ generates a spatially extended actual occasion.

The general process also includes the anticipating component and it accelerates the spatial spreading of an actual occasion. Suppose again that the initial actual occasion $Q(0)$ was pointwise. However, as the result of "feeling" of surrounding actual occasions and processes of creation of new ones, $Q(0)$ can lead to a spatially extended actual occasion.

In the quantum-like model, spatial extension of an information transformer (quantum-like system) is represented in the form of the spatially extended wave function $\Psi(x)$ belonging to the state space $L^2(\mathbf{R}^3)$.[8] From Whitehead's viewpoint, this state presents the information about actual occasions which can be potentially generated in future. Thus *the wave function $\Psi(x)$ is of the anticipatory nature (and it also integrates the memory about preceding actual occasions).* In Philosophy of Organism $\Psi(x)$ cannot be interpreted as the "physical state". In particular, if the wave function $\Psi(x)$ of say an electron is nonzero in two disjoint domains V1 and V2, this does not mean that the electron is "physically present" in both of them. Such $\Psi(x)$ just anticipate the possibility of electron's "occasion" either in V1 or in V2. Such a viewpoint matches perfectly with the information interpretation of quantum mechanics (Zeilinger, 1999, 2010; Brukner and Zeilinger, 1999, 2009; D' Ariano, 2007; Chiribella, D'Ariano, and Perinotti, 2012).

---

[8] A system *S* which is sharply located in the fixed point, say *y*, has to be represented by the wave function $\Psi(x)= \delta(x-y)$. But the latter does not belong to the space of square integrable functions. Thus it cannot be treated as a quantum state. Any nonzero element of $L^2(\mathbf{R}^3)$ has the support on nonzero (Lebesgue) measure.

Similar argument applied to a group of actual occasions, say n occasions $\Phi,..., \Sigma,$ leads to an integral actual occasion which is spatially represented in $\boldsymbol{R}^{3n}$. In quantum mechanics, we operate with the wave function $\Psi(X_1,..., X_n)$, where $X_j$ belongs to $\boldsymbol{R}^3$. Thus an entangled quantum state can be interpreted as representing anticipation of potential realizations of a composite actual occasion.

In short, by the Whiteheadian interpretation quantum nonlocality is *information nonlocality generated by memory about antecedent actual occasions as well as anticipation of potential surrounding occasions.* This interpretation can be compared with the Bohm-Hiley (1993) interpretation of the wave function as the field of active information.

Again in short, the Whiteheadian philosophic system is very supporting for our model of bio-physical evolution – the model of Darwinian natural selection completed by the strong Lamarckian dimension (corresponding to the anticipatory feature of the process of creation of actual occasions).

## Acknowledgments

This work was supported (A. Khrennikov) by the EU-project "Quantum Information Access and Retrieval Theory" (QUARTZ), Grant No. 721321.

**References**

Adami, C., 2003. Sequence complexity in Darwinian evolution. Complexity 8,49–56.

Aerts, S., 2008. An operational characterization for optimal observation of potentialproperties in quantum theory and signal analysis. Int. J. Theor. Phys. 47, 2–14.

Aldrich, H.E., Hodgson, G.M., Hull, D.L., Knudsen, T., Mokyr, J., Vanberg, V.J., 2008. In defence of generalized Darwinism. J. Evol. Econ. 18, 577-596.

Asano, M., Khrennikov, A., Ohya, M., Tanaka, Y., Yamato, I., 2014. Quantum Adaptivity in Biology: from Genetics to Cognition. Springer, Heidelberg-Berlin-New York.

Asano, M., Basieva, I., Khrennikov, A., Ohya, M., Tanaka, Y., Yamato, I., 2015.Quantum information biology: from information interpretation of quantum mechanics to applications in molecular biology and cognitive psychology. Found. Phys. 45, 1362–1378.

Baladrón, C., 2010. In search of the adaptive foundations of quantum mechanics, Physica E 42, 335–338.

Baladrón, C., 2014. Elements for the development of a Darwinian scheme leading to quantum mechanics, in: Nieuwenhuizen, T., et al. (Eds.), Quantum Foundations and Open Quantum systems. World Scientific, Singapore, pp. 489-519.

Baladrón, C., 2016. Physical microscopic free-choice model in the framework of a Darwinian approach to quantum mechanics. Fortschr. Phys..doi:10.1002/prop.201600052

Baladrón, C., Khrennikov, A., 2016. Quantum formalism as an optimisation procedure of informationflows for physical and biological systems. BioSystems 150, 13–21.

D. Barker-Plummer, Turing machines, in ed. Edward N. Zalta, The Stanford Encyclopedia of Philosophy (Winter 2016 Edition). http://plato.stanford.edu/archives/win2016/entries/turing-machine/.

Bennett, C.H., 1988. Logical depth and physical complexity, in: Herken, R. (Ed.), The Universal Turing Machine. A Half-Century Survey. Oxford University Press,Oxford, pp. 227–257.

Bohm D., Hiley B.J., 1993. The Undivided Universe. An Ontological Interpretation of Quantum Theory, Routledge, London.

Brukner, C. and Zeilinger, A.,1999. Malus' law and quantum information. Acta Physica Slovava, 49(4), 647-652.

Brukner, C. and Zeilinger, A., 2009. Information invariance and quantum probabilities. Found. Phys. 39, 677 (2009).

Bub, J., 2016. Quantum Entanglement and Information, in: Zalta, E.N. (Ed.), The Stanford Encyclopedia of Philosophy (Winter 2016 Edition), URL = <https://plato.stanford.edu/archives/win2016/entries/qt-entangle/>.

Chatterjee, K., Pavlogiannis A., Adlam, B., Nowak, M.A., 2013. The time scale ofevolutionary trajectories. <hal-00907940>

Chiribella, G., D'Ariano, G. M. and Perinotti, P., 2012. Informational axioms for quantum theory. In: Foundations of Probability and Physics - 6, D' Ariano, M., Fei, Sh.-M., Haven, E., Hiesmayr, B., Jaeger, G., Khrennikov, A. and J.-A. Larsson (eds.). AIP Conf. Proc. 1424, 270-279.

D' Ariano G M., 2007. Operational axioms for quantum mechanics, in Adenier et al., In: Foundations of Probability and Physics-3, Khrennikov, A. (ed.). AIP Conf. Proc. 889, 79-105.

Dawkins, R., 1983. Universal Darwinism, in: Bendall, D.S. (Ed.), Evolution from Molecules to Man. Cambridge University Press, Cambridge, pp. 403-425.

De Barros, J.A., Suppes, P., 2009. Quantum mechanics, interference, and the brain. J. Math. Psych. 53, 306-313.

De Barros, J.A., 2012. Quantum-like model of behavioral response computation using neural oscillators. Biosystems 110, 171-182.

De Raedt, H., Kastnelson, M.I., Michielsen, K., 2013, Quantum theory as the mostrobust description of reproducible experiments. arXiv:1303.4574 [quant-ph].

De Raedt, H., Katsnelson, M.I., Donkerb, H.C., Michielsen, K., 2015. Quantum theoryas a description of robust experiments: derivation of the Pauli equation. Ann.Phys. 359, 166–186.

Deutsch, D., 1985. Quantum theory, the Church-Turing principle and the universalquantum computer. Proc. R. Soc. Lond. A 400, 97–117.

Eigen, M., 2013. From Strange Simplicity to Complex Familiarity: A Treatise on Matter, Information, Life and Thought. Oxford University Press.

Encyclopedia of Mathematics, 2013. Turing machine. URL: http://www.encyclopediaofmath.org/index.php?title=Turing_machine&oldid=31220 (accessed 29.03.16)

Frank, S.A., 2009. Natural selection maximizes Fisher information. J. Evol. Biol. 22, 231–244.

Frieden, B.R., 1989. Fisher information as the basis for the Schrödinger wave equation. Am. J. Phys. 57, 1004–1008.

Frieden, B.R., Soffer, B.H., 1995. Lagrangians of physics and the game of Fisher information transfer. Phys. Rev. E 52, 2274–2286.

Gernert, D., 2006. Pragmatic Information: Historical Exposition and General Overview. Mind and Matter 4, 141-167.

Goldstein, S., 2016. Bohmian Mechanics, in: Zalta, E.N. (Ed.), The Stanford Encyclopedia of Philosophy (Fall 2016 Edition), URL = <https://plato.stanford.edu/archives/fall2016/entries/qm-bohm/>.

Hameroff, S., 1994. Quantum coherence in microtubules. A neural basis for emergent consciousness? J. Cons. Studies 1, 91-118.

Healey, R., 2016. Quantum-Bayesian and Pragmatist Views of Quantum Theory, in: Zalta, E.N. (Ed.), The Stanford Encyclopedia of Philosophy (Winter 2016 Edition), URL = <https://plato.stanford.edu/archives/win2017/entries/quantum-bayesian/>.

Hiley, B.J., 2002. From the Heisenberg picture to Bohm: A new perspective on active information and its relation to Shannon information. In Quantum Theory: Reconsideration of Foundations, ed. by A. Khrennikov, Växjö University Press, Växjö, pp. 141–162.

Hill, A.L., Nowak, M.A., 2014. Mind over matter. Trends in Ecology and Evolution 29, 74-75.

Honig, J.M., 2009. The role of Fisher information theory in the development of fundamental laws in physical chemistry. Journal of Chemical Education 86, 116-119.

Jaeger, G., 2007. Quantum Information: An Overview. Springer, New York.

Khrennikov, A., 1999.Classical and quantum mechanics on information spaces with applications to cognitive, psychological, social and anomalous phenomena. Found. of Physics, 29, 1065-1098.

Khrennikov, A., 2016. After Bell. Fortschr. Phys..doi:10.1002/prop.201600044

Maynard Smith, J. J., 1974. Theor. Biol. 47, 209-221.

Melkikh, A.V., 2014. Quantum information and the problem of mechanisms of biological evolution. BioSystems. 115, 33-45.

Melkikh, A.V., Khrennikov, A., 2015. Nontrivial quantum and quantum-like effects in biosystems: Unsolved questions and paradoxes. Progress in Biophysics and Molecular Biology. 119. issue 2, 137-161.

Melkikh, A.V., Khrennikov, A., 2016. Quantum-like model of partially directed evolution. Progress in Biophysics and Molecular Biology. 10.1016/j.pbiomolbio.2016.12.005.

Miconi, T., 2008. Evolution and complexity: The double-edged sword. Artif. Life 14, 325-344.

Nielsen, M. A., Chuang, I., 2000. Quantum Computation and Quantum Information. Cambridge University Press.

Nowak, M.A., Sigmund, K., 2004. Science 303, 793-799.

Penrose, R., 1989. The emperor's new mind. Oxford University Press, Oxford.

Penrose, R., 1994. Shadows of the Mind. Oxford University Press, Oxford.

Perrard, S., Labousse, M., Miskin, M., Fort, E., Couder, Y., 2014. Self-organization into quantized eigenstates of a classical wave-driven particle. Nature Com., 5, 3219.

Perrard, S., Fort, E., Couder, Y., 2016. Wave-based Turing machine: time reversal and information erasing. Phys. Rev. Lett., 117, 094502.

Pombo, C., 2010. Observational motives underlying the choice of algebras in physics. Physica E 42, 273–278.

Pombo, C., 2015. Differentiation with stratification: A principle of theoretical physics in the tradition of the memory art. Found. Phys. 45, 1301-1310.

Roederer, J.G., 2016. Pragmatic information in biology and physics. Phil. Trans. R. Soc. A 374: 20150152

Ross, D., 2006. Game theory, in: Zalta, E.N. (Ed.), The Stanford Encyclopedia of Philosophy (Spring 2006 Edition). URL=<http://plato.stanford.edu/archives/spr2006/entries/game-theory/>.

Shimony, A., 1965. Quantum physics and philosophy of Whitehead. Philosophy in America. Cornell Univ. Press, Itaca.

Shor, P.W., 1997. Polynomial-time algorithms for prime factorization and discretelogarithms on a quantum computer. SIAM J. Comput. 26, 1484–1509.

Smolin, L., 2006, The status of cosmological natural selection. arXiv:hep-th/0612185.

Summhammer, J., 1994. Maximum predictive power and the superposition principle. Int. J. Theor. Phys. 33, 171–178.

Summhammer, J., 2007. Quantum theory as efficient representation ofprobabilistic information. arXiv:quant-ph/ 0701181.

Tegmark, M., 2000. Importance of quantum decoherence in brain processes. Phys. Rev. E 61 (4), 4194–4206.

Timpson, C.G., 2006. Philosophical aspects of quantum information theory. arXiv:quant-ph/0611187.

Unger, R.M., Smolin, L., 2015. The Singular Universe and the Reality of Time. Cambridge University Press.

Valentini, A., 2007. Astrophysical and cosmological tests of quantum theory, J. Phys. A: Math. Theor. 40, 3285–3303.

Wheeler, J. A., 1990. Information, physics, quantum: The search for links, in: Zurek, W. H. (Ed.), Complexity, Entropy, and the Physics of Information. Addison-Wesley, Redwood City, California, pp 3-28.

Whitehead, A., N., 1929. Process and reality. Macmillan, New York.

Whitehead, A., N. , 1933. Adventures of ideas. Macmillan, New York.

Wolfram, S., 2002. A New Kind of Science. Wolfram Media.

Wright, S. 1932. The roles of mutation, inbreeding, crossbreeding, and selection in evolution. Proceedings of the Sixth International Congress of Genetics 1, 356-366.

Zeilinger, A. 1999. A foundational principle for quantum mechanics. Found. Phys., 29(4), 631-643.

Zeilinger, A., 2010. Dance of the Photons: From Einstein to Quantum Teleportation. Farrar, Straus and Giroux, New-York.

Zurek, W.H., 2009. Quantum Darwinism. Nat. Phys. 5, 181–188.

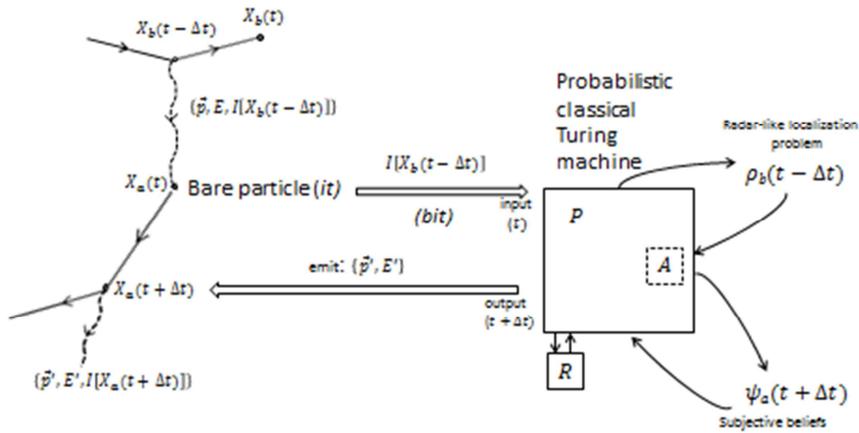

Figure 1: **Representation of a self-interaction cycle of a system**. The information $I[X_b(t - \Delta t)]$ conveyed by the carrier emitted at time $(t - \Delta t)$ by the system $b$ located at $X_b$ and absorbed at $t$ by the system $a$ located at $X_a$ is transferred to the information processor (supplemented with a randomizer $R$) as an input for a run of the program $P$ that calculates, as an intermediate step, the probability distribution function $\rho_b(t - \Delta t)$ of the position occupied by the system $b$ at $(t - \Delta t)$ (and similarly for any surrounding system whose carriers impinge on system $a$). As another intermediate step, and mainly using the anticipation module $A$ of the program, the wave function $\psi_a(t + \Delta t)$ of the system $a$ is computed, reflecting the subjective beliefs of system $a$ about the position to be occupied by the system $b$ at subsequent times (and similarly for any surrounding system whose carriers impinge on the system $a$). As a result, after a convenient post-elaboration of the information and calculated anticipations at disposal (optimization for the stability of the system), a command is generated including the parameters of the carrier to be emitted by the system.